\let\DocumentKernelLabel\label
\documentclass[aps,prl,10pt,amsfonts,amssymb,twocolumn,amsmath,preprintnumbers,nofootinbib,floatfix,superscriptaddress]{revtex4-2}
\let\label\DocumentKernelLabel
\usepackage[T1]{fontenc}
\usepackage{lmodern}   
\usepackage{graphicx}
\usepackage{microtype}
\usepackage{bm}
\usepackage{array}[=2016-10-06]
\usepackage{color}
\usepackage{xurl}
\usepackage[colorlinks,linkcolor=blue,anchorcolor=blue,citecolor=blue,urlcolor=blue]{hyperref}
\usepackage{orcidlink}
\makeatletter
\AtBeginDocument{\let\ltx@label\label}
\let\auto@bib@innerbib\@empty
\let\write@bibliographystyle\relax
\makeatother

\begin{document}
\title{{Flexural Altermagnetism}}

\author{Songbo Li}
\thanks{These authors contributed equally to this work.}
\affiliation{International Center for Quantum Design of Functional Materials,\allowbreak Hefei National Research Center for Physical Sciences at the Microscale, \allowbreak and Department of Physics, \href{https://ror.org/04c4dkn09}{University of Science and Technology of China}, \allowbreak Hefei, Anhui 230026, China}

\author{Jinxiong Jia}
\thanks{These authors contributed equally to this work.}
\affiliation{International Center for Quantum Design of Functional Materials,\allowbreak Hefei National Research Center for Physical Sciences at the Microscale, \allowbreak and Department of Physics, \href{https://ror.org/04c4dkn09}{University of Science and Technology of China}, \allowbreak Hefei, Anhui 230026, China}
\affiliation{Hefei National Laboratory, \href{https://ror.org/04c4dkn09}{University of Science and Technology of China}, \allowbreak Hefei 230088, China}

\author{Zeyu Li}
\thanks{These authors contributed equally to this work.}
\affiliation{Hefei National Laboratory, \href{https://ror.org/04c4dkn09}{University of Science and Technology of China}, \allowbreak Hefei 230088, China}

\author{Xudong Zhu\orcidlink{0000-0003-2006-7109}}
\affiliation{Institute of Artificial Intelligence, Hefei Comprehensive National Science Center, Hefei, 230088, China}

\author{Jiaqi An}
\affiliation{International Center for Quantum Design of Functional Materials,\allowbreak Hefei National Research Center for Physical Sciences at the Microscale, \allowbreak and Department of Physics, \href{https://ror.org/04c4dkn09}{University of Science and Technology of China}, \allowbreak Hefei, Anhui 230026, China}

\author{Feitong~Chu}
\affiliation{International Center for Quantum Design of Functional Materials,\allowbreak Hefei National Research Center for Physical Sciences at the Microscale, \allowbreak and Department of Physics, \href{https://ror.org/04c4dkn09}{University of Science and Technology of China}, \allowbreak Hefei, Anhui 230026, China}

\author{Songbo~Zhang}
\affiliation{International Center for Quantum Design of Functional Materials,\allowbreak Hefei National Research Center for Physical Sciences at the Microscale, \allowbreak and Department of Physics, \href{https://ror.org/04c4dkn09}{University of Science and Technology of China}, \allowbreak Hefei, Anhui 230026, China}
\affiliation{Hefei National Laboratory, \href{https://ror.org/04c4dkn09}{University of Science and Technology of China}, \allowbreak Hefei 230088, China}

\author{Wenguang Zhu}
\affiliation{International Center for Quantum Design of Functional Materials,\allowbreak Hefei National Research Center for Physical Sciences at the Microscale, \allowbreak and Department of Physics, \href{https://ror.org/04c4dkn09}{University of Science and Technology of China}, \allowbreak Hefei, Anhui 230026, China}
\affiliation{Hefei National Laboratory, \href{https://ror.org/04c4dkn09}{University of Science and Technology of China}, \allowbreak Hefei 230088, China}

\author{Hualing Zeng\orcidlink{0000-0001-5869-9553}}
\email{hlzeng@ustc.edu.cn}
\affiliation{International Center for Quantum Design of Functional Materials,\allowbreak Hefei National Research Center for Physical Sciences at the Microscale, \allowbreak and Department of Physics, \href{https://ror.org/04c4dkn09}{University of Science and Technology of China}, \allowbreak Hefei, Anhui 230026, China}
\affiliation{Hefei National Laboratory, \href{https://ror.org/04c4dkn09}{University of Science and Technology of China}, \allowbreak Hefei 230088, China}

\author{Zhenhua Qiao\orcidlink{0000-0002-7195-8561}}
\email{qiao@ustc.edu.cn}
\affiliation{International Center for Quantum Design of Functional Materials,\allowbreak Hefei National Research Center for Physical Sciences at the Microscale, \allowbreak and Department of Physics, \href{https://ror.org/04c4dkn09}{University of Science and Technology of China}, \allowbreak Hefei, Anhui 230026, China}
\affiliation{Hefei National Laboratory, \href{https://ror.org/04c4dkn09}{University of Science and Technology of China}, \allowbreak Hefei 230088, China}
\affiliation{Wilczek Quantum Center, Shanghai Institute for Advanced Studies, \href{https://ror.org/04c4dkn09}{University of Science and Technology of China}, \allowbreak Shanghai 201315, China}

\begin{abstract}
We establish flexural deformation as a versatile route for engineering altermagnetism in two-dimensional collinear antiferromagnets.
By tailoring the flexure, we selectively break different symmetries to realize altermagnetic states with distinct wave symmetries and reverse their spin splitting.
A tight-binding model and first-principles calculations for monolayer MnPS$_3$ demonstrate $d$- and $i$-wave states and a switchable transverse spin current in the $d$-wave state.
\end{abstract}

\maketitle
\noindent{\textit{\textcolor{blue}{Introduction}}}---Altermagnets (AMs) combine compensated collinear magnetic order with momentum-dependent nonrelativistic spin splitting \cite{2022-PRX-Smejkal-Altermagnetism,2022-PRX-Smejkal-Research_landscape,2022-PRX-Liu-Spin_group,2024-PRX-Chen-Spin_space_groups,2022-PRX-Mazin-Altermagnetism_editorial,2024-PRL-Yuan-Brillouin_zone_center,2025-PRL-Liu-Compensated_ferrimagnetism,2020-PRB-Yuan-Spin_splitting}.
This combination offers opportunities for diverse physical phenomena and applications, including spin-polarized transport, large magnetoresistive responses, unconventional superconductivity,
and efficient spin-current generation \cite{2021-PRL-Gonzalez_Hernandez-Spin_splitter,2022-PRX-Smejkal-Magnetoresistance,2024-PRL-Ghorashi-Majorana_modes,2023-PRL-Shao-Neel_spin_currents,2024-PRL-Zhang-Gate_field_control}.
Despite recent experimental progress in AMs, including MnTe and CrSb \cite{2024-PRL-Lee-MnTe,2024-Nature-Krempasky-Kramers_degeneracy,2024-NC-Reimers-CrSb,2024-PRL-Ding-CrSb},
the number of experimentally established AMs remains limited, while some proposed candidates remain under debate \cite{2024-SA-Fedchenko-RuO2,2024-PRL-Liu-RuO2}.
Altermagnetism requires specific spin--space symmetries relating opposite-spin sublattices, which limits the range of suitable materials.
This has motivated efforts to induce altermagnetism in conventional spin-degenerate collinear antiferromagnets (AFMs) \cite{2023-PRL-He-Twisted_bilayers,2024-PRL-Liu-Twisted_bilayers,2025-PRL-Duan-Antiferroelectric_altermagnets,2025-NL-Wang-Electric_field_altermagnets,2025-NQM-Li-Cairo_monolayer,2025-PRB-Ding-Ferroelastic_altermagnets,2025-PRL-Chen-Electrical_switching,2025-PRL-Gu-Ferroelectric_altermagnetism,2025-N2DMA-Gonzalez-Layer_shifts,2021-PNAS-Wu-Sliding_ferroelectricity}.
However, approaches such as twisting require precise control over the interlayer twist angle and are difficult to reconfigure after fabrication,
whereas homogeneous strain generally preserves inversion and translational symmetries, limiting its ability to induce altermagnetism in AFMs protected by these symmetries.
These limitations motivate the exploration of alternative routes to inducing altermagnetism in conventional AFMs.

\begin{figure}[!t]
\centering
\includegraphics[width=\columnwidth]{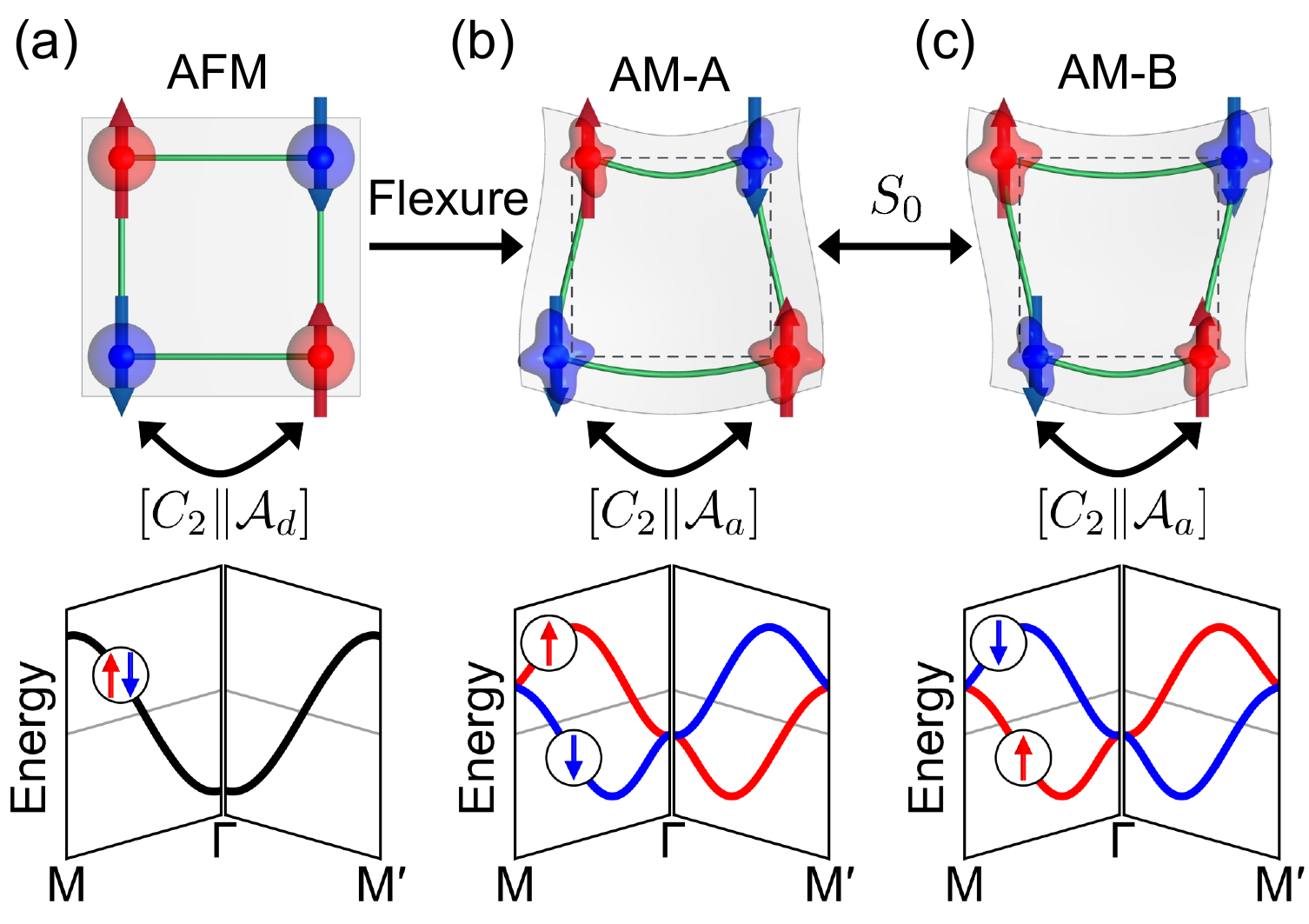}

\vspace{1.5mm}

\begin{minipage}{0.94\columnwidth}
\centering
{\footnotesize
\renewcommand{\arraystretch}{1.5}
\begin{tabular*}{\linewidth}{ @{\hspace{3pt}} c@{\extracolsep{\fill}}cc @{\hspace{3pt}} }
  \hline\hline
  Operation & Momentum mapping & Examples \\
  \hline
  $\mathcal A_d$ & $\mathcal A_d\mathbf{k}=\pm\mathbf{k}\ \forall\mathbf{k}$ & $P,\{E|\boldsymbol{\tau}_a\},M_z,C_{2z}$ \\
  $\mathcal A_a$ & $\mathcal A_a\mathbf{k}\neq\pm\mathbf{k}\ \exists\mathbf{k}$ & $M_v,C_{2\parallel},C_{4z},S_{4z}$ \\
  \hline\hline
\end{tabular*}
}
\end{minipage}

\caption{
Flexural altermagnetism and spin-splitting reversal.
(a) Spin-degenerate antiferromagnet (AFM) state.
(b) and (c) Two symmetry-related flexural altermagnet (AM) configurations with reversed spin splitting.
The upper and lower panels show the real-space spin configurations and spin-resolved bands, respectively.
Red and blue denote opposite spins.
The spin--space operation $S_0$ transforms AM-A into AM-B, reversing the spin splitting.
The table lists representative spin-exchanging spatial operations of two types, $\mathcal A_d$ and $\mathcal A_a$, distinguished by their action on the in-plane momentum.
}
\label{fig:schematic}
\end{figure}

Flexure, characterized by nonuniform out-of-plane lattice deformation, offers a distinct degree of freedom for tuning the spin--space symmetries that relate opposite- spin sublattices.
Owing to their mechanical flexibility, two-dimensional materials commonly develop out-of-plane ripples and wrinkles during growth, transfer, or substrate integration,
making flexural deformation widespread and even unavoidable \cite{2007-Nature-Meyer-Suspended_graphene,2008-Science-Lee-Graphene_strength,2008-PRL-Vazquez_de_Parga-Rippled_graphene,2016-NC-Khestanova-Bubbles,2019-AM-Dai-Strain_engineering}.
Recent experimental advances have also enabled controlled flexural deformations through nano-tip indentation, compliant substrates,
and patterned structures \cite{2024-NC-Liu-Flexoelectric_domains,2020-ACSNano-Rhee-Graphene_wrinkles,2017-NC-Branny-Quantum_emitters,2022-PRApplied-Giambastiani-Graphene_wrinkles}.
These advances motivate us to consider flexure a promising route to activating altermagnetism in spin-degenerate AFMs by controlling spin-exchange symmetries.

In this Letter, based on spin-layer-group symmetry, we illustrate the mechanism for inducing and tuning altermagnetism through flexural deformation.
We show that flexure can break the symmetry-protected global spin degeneracy while preserving those required for altermagnetism, corresponding to a transition from conventional AFMs to AMs.
Notably, when two flexural AM configurations are related by a spin--space symmetry of the parent AFM, opposite-sign AM spin splittings emerge.
Using a generic honeycomb tight-binding model and first-principles calculations for monolayer MnPS$_3$,
we demonstrate that different flexural configurations can induce $d$- and $i$-wave AM states and that symmetry-related configurations exhibit opposite spin splitting.
Finally, we show that the flexure-induced $d$-wave AM state in monolayer MnPS$_3$ supports a transverse spin-splitter response tunable by carrier doping and electric-field orientation and reversible through switching between symmetry-related flexural configurations.

\bigskip
\noindent{\textit{\textcolor{blue}{Symmetry analysis and design principles}}}---Here, we focus on two-dimensional collinear magnets in the absence of spin--orbit coupling (SOC),
where spin and spatial transformations act independently, and their symmetries can be described by collinear spin layer groups (cSLGs) \cite{2026-PRL-Tian-Symmetry_classification,2024-PRX-Chen-Spin_space_groups,2022-PRX-Liu-Spin_group}.
For collinear AFMs and AMs, the relevant discrete symmetry operations, with common spin-only symmetries omitted, can be organized into two cosets,
\begin{equation}
  G_{\mathrm{SLG}} = [E\mathord{\parallel}H] + [C_2\mathord{\parallel}gH],
  \label{eq:spin_layer_group}
\end{equation}
where $E$ is the spin identity and $C_2$ denotes a twofold rotation in spin space that reverses the collinear spins.
Here, $H$ contains spatial operations that preserve each of the two opposite-spin sublattices, while $g$ denotes a spatial operation that exchanges them.
The set $gH$ then contains all such sublattice-exchanging operations.
Combining a spin-exchanging symmetry operation $[C_2\mathord{\parallel}\mathcal A]$, with $\mathcal A\in gH$,
and the spin-preserving effective time-reversal symmetry $[\bar C_2\mathord{\parallel}\mathcal T]$ ~\cite{2022-PRX-Smejkal-Altermagnetism},
we obtain the following relation for the spin-resolved band energies:
\begin{equation}
  E_{ns}(\mathbf{k}) = E_{n\bar{s}}(\pm\mathcal{A}\mathbf{k}),
  \label{eq:spin_symmetry}
\end{equation}
where $n$ is the band index, $s=\uparrow,\downarrow$, $\bar{s}$ denotes the opposite spin,
and $\mathcal{A}\mathbf{k}$ is the in-plane momentum obtained by applying the operator $\mathcal{A}$ to $\mathbf{k}$.
According to Eq.~(\ref{eq:spin_symmetry}), any spin-exchanging spatial operation satisfying $\mathcal{A}\mathbf{k}=\pm\mathbf{k}$ for all momenta enforces global spin degeneracy, $E_{n\uparrow}(\mathbf{k})=E_{n\downarrow}(\mathbf{k})$.
We denote these degeneracy-protecting operations by $\mathcal{A}_d$, and the remaining spin-exchanging operations,
for which $\mathcal{A}_a\mathbf{k}\neq\pm\mathbf{k}$ for at least one momentum, by $\mathcal{A}_a$.
Representative spatial operations of the two types are summarized in the table in Fig.~\ref{fig:schematic}.

Flexure enables selective control of spin-exchanging symmetries through spatially nonuniform out-of-plane lattice deformations.
As illustrated in Figs.~1(b) and 1(c), a suitably designed flexural deformation can break all degeneracy-protecting $[C_2\mathord{\parallel}\mathcal{A}_d]$ symmetries while preserving at least one $[C_2\mathord{\parallel}\mathcal{A}_a]$ symmetry, thereby converting a conventional AFM into an AM.
The remaining spin-exchanging symmetries relate the opposite-spin bands at symmetry-related momenta, allowing momentum-dependent spin splitting.
Defining the spin splitting as $\Delta E_n(\mathbf{k}) =E_{n\uparrow}(\mathbf{k})-E_{n\downarrow}(\mathbf{k})$, Eq.~\eqref{eq:spin_symmetry} gives
\begin{equation}
  \Delta E_n(\mathcal A_a\mathbf{k}) = -\Delta E_n(\mathbf{k}).
  \label{eq:AM_splitting}
\end{equation}
This relation constrains the momentum dependence and symmetry-protected nodes of the spin splitting.
Depending on the surviving spin-exchanging symmetries, flexure may induce $d$-, $g$-, or $i$-wave AM states.

Beyond inducing altermagnetism, flexure can also reverse the AM spin splitting.
Consider two flexural AM configurations $\mathcal{C}_A$ and $\mathcal{C}_B$, each comprising a deformed atomic structure and its spin arrangement.
Let $S_0=[C_2\mathord{\parallel}\mathcal A_d]$ be a spin-exchanging symmetry of the parent AFM,
where $\mathcal A_d$ denotes a degeneracy-protecting spatial operation satisfying $\mathcal A_d\mathbf{k}=\pm\mathbf{k}$.
Although $S_0$ is broken in both flexural configurations, it relates them through $\mathcal{C}_B=S_0\mathcal{C}_A$.
Their spin-resolved band energies satisfy
\begin{equation}
  E_{ns}^{B}(\mathbf{k}) = E_{n\bar{s}}^{A}(\mathbf{k}),
  \label{eq:spin_band_reversal}
\end{equation}
where the correspondence at the same momentum follows from $\mathcal A_d\mathbf{k}=\pm\mathbf{k}$ and same-spin momentum reciprocity.
Consequently, $\Delta E_n^{B}(\mathbf{k}) =-\Delta E_n^{A}(\mathbf{k})$, demonstrating reversed AM spin splitting between the two configurations [Figs.~1(b) and 1(c)].

\begin{figure*}[!tp]
\centering
\includegraphics[width=\textwidth]{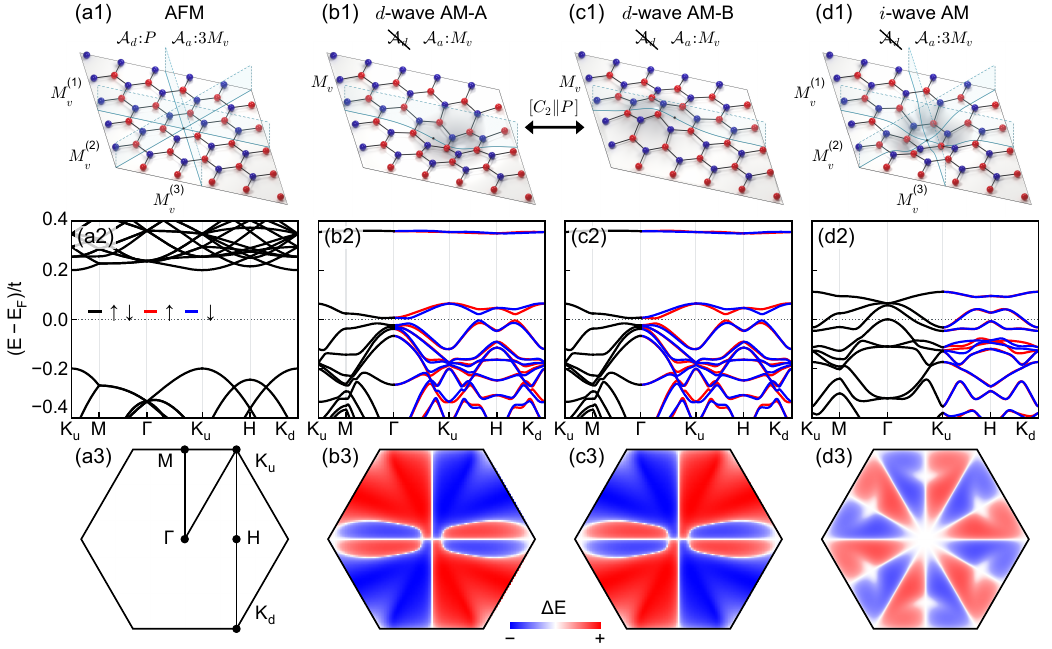}
\caption{
Tight-binding model realization of flexural altermagnetism and spin-splitting reversal.
(a1)--(d1) Lattice configurations of the flat AFM, $d$-wave AM-A, $d$-wave AM-B, and $i$-wave AM, respectively.
(a2)--(d2) Corresponding spin-resolved band structures.
Red and blue denote spin-up and spin-down bands, respectively.
(a3)--(d3) Momentum-resolved spin splitting of the 25th band, $\Delta E(\mathbf{k}) =E_{\uparrow}(\mathbf{k}) -E_{\downarrow}(\mathbf{k})$.
The splitting vanishes for the flat AFM, and (a3) also indicates the Brillouin zone and high-symmetry momentum path.
}
\label{fig:model}
\end{figure*}

\bigskip
\noindent{\textit{\textcolor{blue}{Model}}}---We construct a spin-dependent tight-binding model on a $5\times5$ honeycomb supercell.
In the absence of SOC, the Hamiltonian for spin $s=\pm1$ is
\begin{equation}
  H_s = \sum_{i\neq j}t_{ij}c_{is}^{\dagger}c_{js} + \sum_i (\Phi_i-s\Delta_{\rm ex}\eta_i) c_{is}^{\dagger}c_{is},
  \label{eq:TB_Hamiltonian}
\end{equation}
where $\eta_i=\pm1$ labels the two opposite-spin sublattices and $\Delta_{\rm ex}$ is the staggered exchange field.

We describe the flexural deformation by an out-of-plane height field $h(\mathbf r)$ and a prescribed in-plane displacement field $\mathbf u(\mathbf r)$,
where $\mathbf r$ denotes the atomic position in the undeformed plane.
The deformed atomic positions are given by $\mathbf R(\mathbf r)= [\mathbf r+\mathbf u(\mathbf r),h(\mathbf r)]$.
The resulting changes in bond lengths and local orbital orientations modify the hopping amplitudes.
Assuming that the local $p$-orbital axes follow the surface normals, the Slater--Koster hopping~\cite{1954-PR-Slater-LCAO,2010-PRL-Pereira-Graphene_wrinkles} is
\begin{equation}
  \begin{aligned}
    t_{ij}={}& V_{pp\pi}(d_{ij})\,\mathbf n_i\cdot\mathbf n_j\\
    &+ \left[V_{pp\sigma}(d_{ij}) -V_{pp\pi}(d_{ij})\right] (\mathbf n_i\cdot\hat{\mathbf d}_{ij}) (\mathbf n_j\cdot\hat{\mathbf d}_{ij}),
  \end{aligned}
  \label{eq:SK_hopping}
\end{equation}
where $\mathbf n_i$ is the unit normal to the deformed surface at site $i$, $\mathbf d_{ij}$ is the bond vector between the deformed atomic positions, including periodic images,
$d_{ij}=|\mathbf d_{ij}|$, and $\hat{\mathbf d}_{ij}=\mathbf d_{ij}/d_{ij}$.
We take $V_{pp\pi}(d)=-t e^{-\beta_\pi(d/a-1)}$ and $V_{pp\sigma}(d)= \gamma_\sigma t e^{-\beta_\sigma(d/a-1)}$.
We also include a curvature-dependent onsite potential $\Phi_i=-\lambda_\phi [\nabla^2h(\mathbf r_i)]^2$~\cite{2008-EPL-Kim-Electronic_membrane}.
The deformation profiles and implementation details are given in the Supplemental Material (SM)~\cite{SM}.
Throughout Fig.~2, we use $\Delta_{\rm ex}=0.2t$, $\beta_\pi=\beta_\sigma=1$, $\gamma_\sigma=3.5$, and $\lambda_\phi=12ta^2$, with $a=t=1$.

We first consider the flat AFM configuration, which preserves a spin-exchanging inversion operation $P$ and three vertical mirror symmetries, all exchanging the opposite-spin sublattices [Fig.~2(a1)].
As established above, $[C_2\mathord{\parallel}P]$ protects global spin degeneracy, resulting in coincident spin-up and spin-down bands [Fig.~2(a2)].
Introducing a Gaussian depression offset from the supercell center along a spin-exchanging mirror axis breaks $P$ and two of the three vertical mirror symmetries,
while preserving the mirror symmetry $M_v$ along the displacement direction.
The resulting configuration, denoted AM-A, exhibits $d$-wave AM spin splitting [Fig.~2(b2)].
According to Eq.~\eqref{eq:spin_symmetry}, the preserved symmetry enforces two perpendicular spin-degenerate nodal lines through $\Gamma$, resulting in a $d$-wave AM spin-splitting pattern [Fig.~2(b3)].

To demonstrate spin-splitting reversal, we construct a second flexural configuration AM-B related to AM-A by a spin--space symmetry $S_0=[C_2\mathord{\parallel}P]$ of the parent AFM, $\mathcal C_B=S_0\mathcal C_A$.
As predicted by Eq.~\eqref{eq:spin_band_reversal}, the opposite-spin bands are interchanged between the two configurations [Figs.~2(b2) and 2(c2)].
Consequently, $\Delta E_{25}^{B}(\mathbf k)= -\Delta E_{25}^{A}(\mathbf k)$.
The two configurations retain the same $d$-wave nodal structure but exhibit opposite spin-splitting patterns [Figs.~2(b3) and 2(c3)].

A different flexural geometry realizes $i$-wave altermagnetism.
By arranging three identical Gaussian depressions at $120^\circ$ intervals around the supercell center,
the deformation removes the global-degeneracy-protecting symmetries while retaining three vertical mirror symmetries [Fig.~2(d1)].
The resulting spin splitting exhibits six nodal lines and twelve alternating-sign sectors near $\Gamma$, characteristic of an $i$-wave state [Figs.~2(d2) and 2(d3)].

These results demonstrate that flexure can both induce distinct AM spin-splitting patterns and reverse the spin splitting between symmetry-related configurations.

\begin{figure*}[!tp]
\centering
\includegraphics[width=\textwidth]{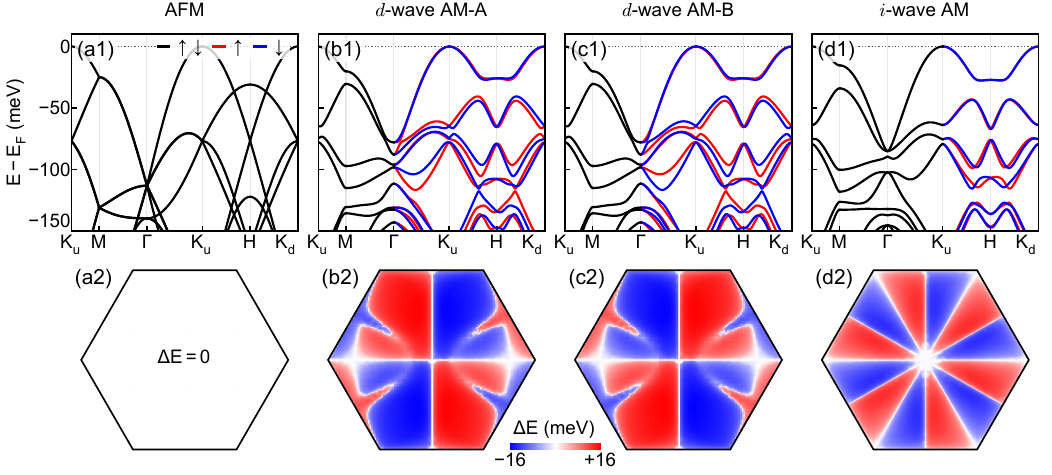}
\caption{
First-principles realization of flexural altermagnetism and spin-splitting reversal in monolayer MnPS$_3$.
(a1)--(d1) Spin-resolved band structures of the flat AFM, $d$-wave AM-A, $d$-wave AM-B, and $i$-wave AM, respectively.
Red and blue denote spin-up and spin-down bands, respectively.
(a2)--(d2) Corresponding momentum-resolved spin splitting of the fourth-highest occupied band, $\Delta E(\mathbf{k})= E_{\uparrow}(\mathbf{k})-E_{\downarrow}(\mathbf{k})$.
}
\label{fig:MnPS3}
\end{figure*}

\bigskip
\noindent{\textit{\textcolor{blue}{First-principles realization in MnPS$_3$}}}---To validate the proposed mechanism in a realistic material,
we perform spin-polarized density-functional calculations for monolayer MnPS$_3$ using the Perdew--Burke--Ernzerhof exchange--correlation functional with a Hubbard-$U$ correction,
as implemented in ABACUS \cite{1996-PRL-Perdew-PBE,1998-PRB-Dudarev-LSDA_U,2022-JCTC-Qu-DFT_DMFT,2016-CMS-Li-Atomic_basis,2025-JCP-Zhou-ABACUS,2022-JPCC-Autieri-MPS3,2023-NCM-Gao-MnPS3_photoexcitation}.
We employ a $4\times4$ supercell with collinear N\'eel order and examine the electronic structures of different flexural configurations.
Computational details are provided in the SM~\cite{SM}.

The flat AFM exhibits global spin degeneracy protected by the spin-exchanging inversion symmetry $P$ [Figs.~\ref{fig:MnPS3}(a1) and \ref{fig:MnPS3}(a2)].
To induce $d$-wave altermagnetism, we introduce an off-center Gaussian-shaped depression along a spin-exchanging mirror axis, followed by constrained structural relaxation.
The resulting structure breaks $P$ while preserving a single spin-exchanging mirror $M_v$, yielding $d$-wave spin splitting [Figs.~\ref{fig:MnPS3}(b1) and \ref{fig:MnPS3}(b2)].
Applying the parent-AFM symmetry $S_0=[C_2\parallel P]$ to this configuration produces AM-B, whose spin splitting is reversed relative to AM-A,
as predicted by Eq.~\eqref{eq:spin_band_reversal} [Figs.~\ref{fig:MnPS3}(c1) and \ref{fig:MnPS3}(c2)].
An $i$-wave state is realized by placing the Gaussian-shaped depression at the supercell center, followed by constrained structural relaxation.
The resulting structure breaks the spin-exchanging inversion symmetry $P$ while retaining three spin-exchanging vertical mirrors,
yielding the characteristic $i$-wave spin-splitting pattern [Figs.~\ref{fig:MnPS3}(d1) and \ref{fig:MnPS3}(d2)].

These results confirm that the symmetry-based mechanism established in the tight-binding model remains effective in a realistic multiorbital system after structural relaxation.

\begin{figure}[!tbp]
\centering
\includegraphics[width=\columnwidth]{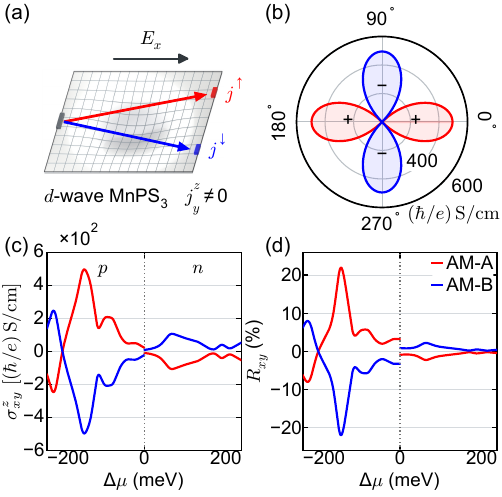}
\caption{
Flexure-controlled spin-current response in monolayer MnPS$_3$.
(a) Schematic of the nonrelativistic spin-splitter effect in the $d$-wave AM, where an in-plane electric field $E_x$ generates opposite transverse spin-channel currents.
(b) Angular dependence of the transverse spin conductivity $\sigma_\perp^z$ at $\Delta\mu_p=-125$~meV.
Red and blue indicate positive and negative responses, respectively.
(c) Transverse spin conductivity $\sigma_{xy}^{z}$ and (d) spin-splitter ratio $R_{xy}$ as functions of the chemical-potential shift $\Delta\mu$ for AM-A (red) and AM-B (blue).
Negative (positive) $\Delta\mu$ corresponds to $p$-type ($n$-type) doping.
}
\label{fig:spin_current}
\end{figure}

\bigskip
\noindent{\textit{\textcolor{blue}{Flexure-controlled spin-current response}}}---Having established flexural altermagnetism in monolayer MnPS$_3$,
we further investigate its transverse spin-current response using the first-principles electronic structures.
Under the relaxation-time approximation, the spin-resolved conductivity is given by \cite{2021-PRL-Gonzalez_Hernandez-Spin_splitter,2025-PRB-Dou-Spin_conductivity,2026-PRB-Sheoran-Spin_currents}
\begin{equation}
  \sigma_{ij}^{s}(\mu)=e^2\tau\sum_n \int_{\mathrm{BZ}}\!\frac{d^2k}{(2\pi)^2} v_{ni}^{s}v_{nj}^{s} \left(-\frac{\partial f_0}{\partial E_{n\mathbf{k}}^{s}}\right),
  \label{eq:charge_conductivity}
\end{equation}
where $v_{ni}^{s}=\hbar^{-1} \partial E_{n\mathbf{k}}^{s}/\partial k_i$ is the band velocity, $\tau$ is the relaxation time, and $f_0$ is the Fermi--Dirac distribution.
The total charge conductivity, spin conductivity, and spin-splitter ratio are defined as
\begin{equation}
  \begin{aligned}
    \sigma_{ij}^{c} &=\sigma_{ij}^{\uparrow} +\sigma_{ij}^{\downarrow},\\
    \sigma_{ij}^{z} &=\frac{\hbar}{2e} (\sigma_{ij}^{\uparrow} -\sigma_{ij}^{\downarrow}),\\
    R_{xy} &=\frac{2e}{\hbar} \frac{\sigma_{xy}^{z}}{\sigma_{xx}^{c}}.
  \end{aligned}
  \label{eq:spin_response}
\end{equation}
Here $\sigma_{ij}^{z}$ denotes the spin conductivity polarized along $z$, while $R_{xy}$ characterizes the transverse spin-current response relative to the longitudinal charge current.
Within this transport approximation, $\sigma_{xy}^{z}=\sigma_{yx}^{z}$, so an electric field along $x$ generates $j_y^z=\sigma_{yx}^{z}E_x$,
corresponding to the nonrelativistic spin-splitter effect illustrated in Fig.~\ref{fig:spin_current}(a).

For the $d$-wave AM state of monolayer MnPS$_3$, we examine the angular dependence of the transverse spin conductivity.
For an in-plane electric field $\mathbf{E}=E(\cos\theta,\sin\theta)$, the transverse spin conductivity is defined as $\sigma_\perp^z=(j_y^z\cos\theta-j_x^z\sin\theta)/E$.
As shown in Fig.~\ref{fig:spin_current}(b), $\sigma_\perp^z$ reverses sign upon a $90^\circ$ rotation of the electric field at $\Delta\mu_p=-125$~meV, consistent with the $d$-wave symmetry.

More importantly, the AM-A and AM-B configurations of monolayer MnPS$_3$ are related by symmetry $[C_2\mathord{\parallel} P]$, which reverses their transverse spin responses,
yielding $\sigma_{xy}^{z,B}=-\sigma_{xy}^{z,A}$ and $R_{xy}^{B}=-R_{xy}^{A}$.
Figures~\ref{fig:spin_current}(c) and \ref{fig:spin_current}(d) show the chemical-potential dependence of $\sigma_{xy}^{z}$ and $R_{xy}$,
respectively, confirming the symmetry-enforced reversal between AM-A and AM-B.
For a broadening of $\Gamma=10$~meV, $|R_{xy}|$ reaches approximately $20\%$ on the hole-doped side near $\Delta\mu_p=-125$~meV.
These results demonstrate that switching between symmetry-related flexural configurations can reverse the transverse spin-current response,
providing mechanical control of the nonrelativistic spin-splitter effect.
The transport method and slab normalization are given in SM~\cite{SM}.

\bigskip
\noindent{\textit{\textcolor{blue}{Conclusion and outlook}}}---In this work, we establish flexure as a mechanical degree of freedom for inducing and controlling altermagnetism in two-dimensional AFMs.
Our spin-layer-group analysis reveals that flexural deformation can selectively break the symmetries protecting global spin degeneracy while preserving those required for altermagnetism.
A honeycomb tight-binding model and first-principles calculations for monolayer MnPS$_3$ demonstrate distinct $d$- and $i$-wave AM states and the reversal of spin splitting between symmetry-related flexural configurations.
Furthermore, switching between these configurations reverses the transverse spin-current response, demonstrating mechanical control of the nonrelativistic spin-splitter effect.

Flexure can also induce ferrimagnetism, as discussed in the SM~\cite{SM}.

Flexure offers a promising route to inducing altermagnetism in conventional spin-degenerate AFMs and reversibly controlling spin splitting,
thereby broadening the range of candidate materials and opening opportunities for spin-current generation and switching.
In two-dimensional materials, flexural deformations can arise naturally as ripples and wrinkles, potentially giving rise to altermagnetism when the required symmetry conditions are satisfied.
Such deformations can also be deliberately introduced and controlled through nano-tip indentation, compliant substrates, and mechanical actuators,
providing practical pathways toward reconfigurable spintronic devices.
Although our study focuses on two-dimensional systems, the underlying symmetry-based principle may also extend to three-dimensional AFMs,
opening broader possibilities for mechanically programmable spintronics.
\par

\bigskip
\noindent{\bfseries Acknowledgments} ---This work was supported by the National Key R\&D Program of China (Grant No. 2024YFA1408103),
the National Natural Science Foundation of China (Grants No. 12474158, 12234017, 12488101, 12404288 and 12504195),
the Quantum Science and Technology-National Science and Technology Major Project (Grant No. 2021ZD0302801), and the Fundamental Research Funds for the Central Universities (Grant No. WK9990000132).
We also thank the Supercomputing Center of University of Science and Technology of China for providing high-performance computing resources.

\end{document}